\documentclass[aps,prc,showpacs,nofootinbib]{revtex4}

\usepackage{epsfig}
\usepackage{graphicx}

\begin{document}

\title{Polarization phenomena in elastic $e^{\mp}N$ scattering, for axial parametrization of two-photon exchange}

\author{M. P. Rekalo}
\affiliation{
\it NSC Kharkov Physical Technical Institute, 61108 Kharkov, Ukraine}

\author{E. Tomasi-Gustafsson}
\affiliation{\it DAPNIA/SPhN, CEA/Saclay, 91191 Gif-sur-Yvette Cedex,
France }

\date{\today}
\pacs{25.30.Bf, 13.40.-f, 13.60.-Hb, 13.88.+e}

\begin{abstract}
We analyze polarization phenomena for elastic lepton-nucleon scattering, parametrizing the $2\gamma$-exchange contribution, as the product of lepton and nucleon axial vector currents, that correspond to $C=+1$ exchange in the annihilation channel. We found two combinations of polarization observables (one for three T-odd observables and another one - for five T-even observables), which allow to measure the ratio  $G_E(Q^2)/G_M(Q^2)$ of nucleon electromagnetic form factors in model independent way, without any specific assumption about the $2\gamma$-exchange mechanism. Both these combinations have a general character and do not depend on the choice of the spin structure for the $2\gamma$-exchange. We show the inconsistency of other approximations recently used in the literature.

\end{abstract}

\maketitle
\section{Introduction}

The presence of $2\gamma$-exchange in elastic $e^{\mp}N$ scattering, can essentially change the traditional procedure of extracting information about the nucleon electromagnetic structure \cite{Ho62}. The necessity to introduce an additional amplitude to describe this exchange, leads to a complicated analysis of polarization effects in elastic and inelastic $eN$-scattering (as well as in other processes of elastic and inelastic electron scattering by different hadronic targets) in order to determine the corresponding electromagnetic form factors \cite{Re03a,Re03b}. This is not only due to the presence of an additional amplitude, but mainly to the fact that all amplitudes for elastic $eN$-scattering must be complex functions of both kinematical variables, in the presence of $2\gamma$-contributions.

We proved earlier \cite{Re03b}, that, in any case, the realization of the complete experiment for elastic $eN$-scattering, (which can be done in at least two possible ways) allows to measure the ratio $G_E(Q^2)/G_M(Q^2)$ of physical nucleon electromagnetic FFs. Both methods differ essentially from the standard procedure for the determination of the FFs \cite{Ro50,Re68,Re03l}. In particular, the $2\gamma$ contribution destroys the linearity in the variable $\epsilon$ (the degree of polarization of the virtual photon) of the differential cross section for elastic $eN$ scattering, and the relatively simple dependence of the ratio $P_x/P_z$ (the components of the final nucleon polarization in the scattering of longitudinally polarized electrons by an unpolarized nucleon target)  on the ratio $G_E(Q^2)/G_M(Q^2)$, which can be predicted for the one-photon mechanism \cite{Re68,Re03l}.

However, both the above mentioned methods  involve the measurement of very small effects ($\simeq\alpha=1/137$) of T-odd nature, or the measurement of relatively large T-even observables, but with very large accuracy, i.e. better than 1\%. Note that, in any case, such complementary procedure seems unavoidable, in presence of the $2\gamma$ contribution.

In framework of the formalism, necessary to describe the $2\gamma$-contribution, 'generalized' FFs have been introduced, $\widetilde{G}_{E,M}$, which are complex functions of both kinematical variables $Q^2$ and $\epsilon$ \cite{Gu03}. Such functions do not have a straightforward physical interpretation, as they can not be uniquely defined. Therefore, the experimental procedure to determine these 'generalized' FFs depends on the parametrization of the additional $2\gamma$ spin structure, for the matrix element in  $eN$-scattering. Therefore, different parametrizations lead to different sets of polarization experiments for the determination of $G_{E,M}(Q^2,\epsilon)$. This means that the 'generalized' FFs are not physical observables, and instead of the simple coherent and unique formalism to determine the real FFs, $G_{E,M}(Q^2)$, one has to deal with non-physical quantities. As a result, any procedure to simplify the general analysis of processes in presence of $2\gamma$ exchange, through additional and unjustified assumptions, will give different answers, depending on the parametrization of the general matrix element. This explains the disagreement of recent analysis of existing experimental data about elastic $eN$-scattering \cite{Ar03}.

The aim of this paper is to continue our previous analysis of polarization phenomena for elastic $eN$ scattering, in presence of the $2\gamma$-contribution, using a special parametrization of the spin structure of the matrix element. Such parametrization has been suggested long ago, \cite{Drell}, and describes the $2\gamma$ exchange as an axial meson exchange, with positive C-parity. This is a simplified picture, because this model generates a $2\gamma$-real amplitudes, proportional to the axial nucleon form factor.

But the corresponding spin structure, as a product of electron and nucleon axial currents, has a more general character, and can therefore be used for the description of the $2\gamma$-contribution. This means that a new, additional invariant amplitude has to be introduced here, which in general, is complex function of both independent kinematical variables, $A=A(Q^2,\epsilon)$. Such parametrization, henceforth  called the axial parametrization, has to be completely equivalent to the parametrization from \cite{Re03b}, henceforth  called the tensor parametrization.

We derive here polarization phenomena for elastic $eN$-scattering, using the axial parametrization of the $2\gamma$ contribution. The expressions for the polarization observables will differ from the ones derived in \cite{Re03b}, but the final strategy for the measurement of the physical FFs $G_{E,M}(Q^2)$ must be the same.

The paper is organized as follows. After a discussion of the general parametrization of the matrix element for elastic $eN$ scattering, using the axial form (section II), we present, in Section III, the corresponding formulas for the differential cross section and for the $P_x$ and $P_z$ components of the final nucleon polarization, in terms of three invariant amplitudes. These formulas allow us to prove that the measurement of these three observables in electron and positron scattering can be considered the simplest way for the determination of  $G_{E,M}(Q^2)$. T-odd polarization effects, analyzed in Section IV, give a specific combination of observables, which is proportional to $G_E(Q^2)/G_M(Q^2)$. An equivalent combination of T-even observables is derived in Section V.
%%%%%%%%%%%%%%%%%%%%%%%%%%%%%%%%%%%%%%%%%%%%%%%%%%%%%%%%%%%%%%%%%%
\section{The matrix element for $eN$ elastic scattering: axial parametrization}
%%%%%%%%%%%%%%%%%%%%%%%%%%%%%%%%%%%%%%%%%%%%%%%%%%%%%%%%%%%%%%%%%%
The general structure of the matrix element for $eN$ elastic scattering, which can be written for the one-photon + two-photon exchange mechanism is
 \cite{Drell}:
\begin{eqnarray}
{\cal  M}^{(\mp)}&=&\displaystyle\frac{e^2}{Q^2}\left \{\overline{u}(k_2)\gamma_{\mu}u(k_1)
\overline{u}(p_2)  \left [ \gamma_{\mu} F_1^{(\mp)}(Q^2,\epsilon) +\right .
\displaystyle\frac{\sigma_{\mu\nu}q_{\nu}}{2m}F_2^{(\mp)}(Q^2,\epsilon)\right ] u(p_1)
\nonumber\\
&&\left .+\overline{u}(k_2)\gamma_{\mu}\gamma_5u(k_1)\overline{u}(p_2)\gamma_{\mu}\gamma_5u(p_1){\cal  A}^{(\mp)}(Q^2,\epsilon)\right \}, 
\label{eq:mat1}
\end{eqnarray}
where  $k_1$ $(p_1)$ and $k_2$ $(p_2)$ are the four-momenta of the initial and final electron (nucleon), $m$ is the nucleon mass, $q=k_1-k_2$, $Q^2=-q^2>0$, and the upper index $(\mp)$ corresponds to $e^-(e^+)$ scattering. The variable $\epsilon$ characterizes the virtual photon polarization (in the case of the one-photon mechanism):
\begin{equation}
\epsilon=
\displaystyle\frac{1}{1-2\displaystyle\frac{\vec q^2}{Q^2}\tan^2 \displaystyle\frac{\theta_e}{2}},
\label{eq:csst1}
\end{equation}
where $\theta_e$ is the electron scattering angle in Lab system, $\vec q$ is the three momentum transfer. Therefore $0 (\theta_e=\pi)\le \epsilon\le 1 (\theta_e=0)$. 

The three complex amplitudes,$ F_{1,2}^{(\mp)}(Q^2,\epsilon)$ and ${\cal  A}^{(\mp)}(Q^2,\epsilon)$, which, generally are functions of two independent kinematical variables, $ Q^2$ and $\epsilon$, fully describe the spin structure of the matrix elements for $ e^{\mp}N$ elastic scattering - for any number of exchanged photons.

This expression holds under assumption of P-invariant electromagnetic interaction and conservation of hadron helicity, which is correct for standard QED at high energy, i.e. in the limit of zero electron mass. Note, however, that  Eq. (\ref{eq:mat1}) is one of the many equivalent representations of the matrix element for $ e^{\mp}N$ elastic scattering.

The C-invariance and the crossing symmetry of the $eN$-interaction allows to write (in the $1\gamma$ +  $2\gamma$ approximation):
\begin{eqnarray}
F_{1,2}^{(\mp)}(Q^2,\epsilon)&=&\mp F_{1,2}(Q^2)+\Delta F_{1,2}^{(\mp)}(Q^2,\epsilon),\nonumber\\
\Delta F_{1,2}^{(-)}(Q^2,\epsilon)&=&\Delta F_{1,2}^{(+)}(Q^2,\epsilon)\equiv \Delta F_{1,2}(Q^2,\epsilon),
\label{eq:eq2}\\
{\cal  A}^{(-)}(Q^2,\epsilon)&=&{\cal  A}^{(+)}(Q^2,\epsilon)\equiv {\cal  A}(Q^2,\epsilon),
\nonumber
\end{eqnarray}
where $F_{1,2}(Q^2)$ are the Dirac and Pauli nucleon electromagnetic FFs, which are real function of one variable, $Q^2$, only (in the space-like region of momentum transfer square). It follows from (\ref{eq:eq2}), that both elastic processes, $e^{\mp}+N\to e^{\mp}+N$, with six complex amplitudes, 
$F_{1,2}^{(\mp)}(Q^2,\epsilon)$, ${\cal  A}^{(\mp)}(Q^2,\epsilon)$ can be described by eight different real functions:
$$ F_{1,2}(Q^2),~Re \Delta F_{1,2}(Q^2,\epsilon) ,~Re{\cal  A}(Q^2,\epsilon),~Im \Delta F_{1,2}(Q^2,\epsilon), \mbox{~and~} Im {\cal  A}(Q^2,\epsilon).$$
The order of magnitude of these quantities is
$$\Delta F_{1,2}(Q^2,\epsilon)\mbox{~and~}{\cal  A}(Q^2,\epsilon)\simeq \alpha\simeq\displaystyle\frac{e^2}{4\pi}\simeq \displaystyle\frac{1}{137},\mbox{~and~} F_{1,2}(Q^2)\simeq \alpha^0.$$
It is important to stress that the main contributions, $F_{1,2}(Q^2)$, are functions of $Q^2$ only.
Note that, due to C-invariance and crossing symmetry, the $\epsilon$ dependence of the other six functions occurs through a specific argument, 
\begin{equation}
x=\sqrt{\displaystyle\frac{1+\epsilon}{1-\epsilon}},
\label{eq:eqxx}
\end{equation}
with the following symmetry properties, with respect to the change $x\to -x$:
\begin{eqnarray}
\Delta F_{1,2}(Q^2,-x)&=& -\Delta F_{1,2}(Q^2,x),\label{eq:eq3a}\\
{\cal A}_3(Q^2,-x)&=+&{\cal A}(Q^2,x).\label{eq:eq3b}
\end{eqnarray}

The functions, $\Delta F_{1,2}(Q^2,\epsilon)$ and ${\cal A}(Q^2,\epsilon)$ describe the $2\gamma$ contribution to  the matrix element. Having the axial-vector, i.e. $\gamma_{\mu}\gamma_5$-structure, the corresponding contribution to the matrix element, characterized by the complex amplitude ${\cal  A}(Q^2,\epsilon)$ can not be identified with the nucleon axial FF, $f_A(Q^2)$, which is a real function of $Q^2$, only. In principle, it can be defined again as a 'generalized' axial FF, but this will be source of confusion. The amplitude ${\cal  A}(Q^2,\epsilon)$ describes only the part of the $2\gamma$ contribution, the spin structure of which,  $\gamma_{\mu}\gamma_5$, corresponds to $C=+1$ exchange in the $t$ channel, with definite symmetry properties with respect to $x\to -x$ transformation, Eq. (\ref{eq:eq3b}). The functions $\Delta F_{1,2}(Q^2,\epsilon)$ also characterize the $2\gamma$ contribution, but with other symmetry properties, see Eq. (\ref{eq:eq3a}), due to the vector-like ($C=-1$) nucleonic currents, $\gamma_{\mu}$ and  $\sigma_{\mu\nu}q_{\nu}$.
The choice of these vector structures looks unique, as they appear also for one-photon exchange, whereas the third ($2\gamma$) spin structure can be chosen in different but equivalent ways.

The resulting expressions for the differential cross section and different polarization observables will depend on this choice. But the procedure to extract the physical FFs, $F_{1,2}(Q^2)$ from polarization experiments, can not depend on this choice. The proof is given in this paper, where the expressions for the polarization observables will be derived using the axial parametrization, and compared with the (more complicated) expressions found in the analysis of the complete experiment based on the tensor parametrization \cite{Re03b}. A rigorous analysis of polarization effects has to give the same physical result, concerning the physical quantities, such as the nucleon form factors, independently of the parametrization used (tensor or axial). It has been suggested  \cite{Gu03,Bl03}, that the presence of $2\gamma$ exchange could solve the discrepancy between experimental data on the proton electromagnetic FFs, extracted by different procedures, through the Rosenbluth fit \cite{Ar75}, or from the polarization transfer method  \cite{Re68,Re03l}. Quantitative estimations depend on the choice of the third spin structure. More precisely, additional assumptions have to be done, which translate differently in the axial or tensor polarization. In both cases these assumptions can not be justified on the basis of general principles, but such considerations can solve the contradiction between different sets of data. Unfortunately, it is not a rigorous and unique solution, as some of these simplifying assumptions contradict C-invariance and crossing symmetry. This is true using both parametrizations of the $2\gamma$ term.

\section{Differential cross section and $P_{x,z}$ components of the nucleon polarization }

We discuss here the three observables, the differential cross section $d\sigma/d\Omega_e$, and the $P_x$ and $P_z$ components of the nucleon polarization, in the scattering of longitudinally polarized electrons by unpolarized target, $\vec e^-+N\to e^-+\vec N$, which are the relevant observables for the extraction of the nucleon FFs.

Using the general parametrization of the spin structure of the matrix element, Eq. (\ref{eq:mat1}), one can find the following expressions for these observables, taking into account the main ($\alpha^2$) and the $1\gamma$ + $2\gamma$ contribution, ($\alpha^3$):
\begin{eqnarray}
\displaystyle\frac{d\sigma^{(\mp)}}{d\Omega_e}&=&\sigma_0\left [ \tau |G_M^{(\mp)}(Q^2,\epsilon)|^2+
 \epsilon|G_E^{(\mp)}(Q^2,\epsilon)|^2+ \right .\nonumber \\
&&\hspace*{1true cm} \left .\sqrt{1-\epsilon^2}\sqrt{\tau(1+\tau)}Re G_M^{(\mp)}(Q^2,\epsilon){\cal A}^{(\mp)*}(Q^2,\epsilon)\right ]
\equiv \sigma_0{\cal N}^{(\mp)},
\label{eq:eq4a}
\\
{\cal N}^{(\mp)}P_x^{(\mp)})&=&-\lambda_e\sqrt{2\epsilon(1-\epsilon)\tau}
\left [Re G_E^{(\mp)}(Q^2,\epsilon)G_M^{(\mp)*}(Q^2,\epsilon)+ 
\right .\nonumber\\&&\hspace*{1true cm}\left .
\sqrt{\displaystyle\frac{1+\epsilon}{1-\epsilon}} Re 
G_E^{(\mp)}(Q^2,\epsilon){\cal A}^{(\mp)*}(Q^2,\epsilon)\right ],\label{eq:eq4b}
\\
{\cal N}^{(\mp)}P_z^{(\mp)})&=&\lambda_e\tau\sqrt{1-\epsilon^2}\left [ |G_M^{(\mp)}(Q^2,\epsilon)|^2+ 2 
\sqrt{\displaystyle\frac{(1+\tau)}{\tau(1-\epsilon^2)} } Re G_M^{(\mp)}(Q^2,\epsilon)
{\cal A}^{(\mp)*}(Q^2,\epsilon)|^2\right ], \label{eq:eq4c}
\end{eqnarray}
where $\lambda_e$ is the degree of the initial lepton longitudinal polarization, and 
\begin{eqnarray}
G_M^{(\mp)}(Q^2,\epsilon)=F_1^{(\mp)}(Q^2,\epsilon)+F_2^{(\mp)}(Q^2,\epsilon),\nonumber\\
G_E^{(\mp)}(Q^2,\epsilon)=F_1^{(\mp)}(Q^2,\epsilon)+\tau F_2^{(\mp)}(Q^2,\epsilon).
\label{eq:eq5}
\end{eqnarray}
A quick inspection of Eqs. (\ref{eq:eq4a},\ref{eq:eq4b},\ref{eq:eq4c}) shows that even the complete measurement of the 
$Q^2$ and $\epsilon$ dependencies of these three observables, $(d\sigma/d\Omega_e)^{(-)}$,  $P_x^{(-)}$ and $P_z^{(-)}$ do not allow to find five quadratic combinations of the corresponding amplitudes:
$$
|G_{E,M}^{(-)}(Q^2,\epsilon)|^2,~ Re  G_E^{(-)}(Q^2,\epsilon)~G_M^{(-)*}(Q^2,\epsilon), 
$$
\begin{equation}
{\cal R}_E^{(-)}(Q^2,\epsilon)\equiv Re  G_E^{(-)}(Q^2,\epsilon)
{\cal A}^{(-)*}(Q^2,\epsilon),~{\cal R}_M^{(-)}(Q^2,\epsilon)=Re  G_M^{(-)}(Q^2,\epsilon)
{\cal A}^{(-)*}(Q^2,\epsilon),
\label{eq:eq6}
\end{equation}
which determine these observables. So, additional assumptions have to be done,   to reduce the number of five amplitudes to three, which imply different speculative approaches. This can be done in different ways and the results will be correspondingly different. We will use another procedure, which allows to avoid this simplification.
Using Eq. (\ref{eq:eq2}), we rewrite Eqs. (\ref{eq:eq4a},\ref{eq:eq4b},\ref{eq:eq4c})  in the following form, keeping the $1\gamma\bigotimes 2\gamma$ interference contribution:
\begin{eqnarray}
{\cal N}^{(\mp)}&=&\epsilon \left [ |G_E(Q^2)|^2\mp 2 G_E(Q^2) Re \Delta G_E(Q^2, \epsilon)\right ]+
\nonumber\\
&&\tau \left [ |G_M(Q^2)|^2\mp 2 G_M(Q^2) Re \Delta G_M(Q^2, \epsilon)\right ]\mp \label{eq:eq7a}\\
&&\sqrt{(1-\epsilon^2)\tau(1+\tau)}G_M(Q^2) Re {\cal A}(Q^2,\epsilon),\nonumber\\
{\cal N}^{(\mp)}P_x^{(\mp)})&=&-\lambda_e\sqrt{2\epsilon(1-\epsilon)\tau}
\left \{ G_E(Q^2)G_M(Q^2)\mp Re 
\left [ G_E(Q^2) \Delta G_M(Q^2, \epsilon)\right .+ \right .\nonumber\\
&&\hspace*{1true cm}\left .\left . G_M(Q^2)\Delta G_E(Q^2, \epsilon)+ \sqrt{\displaystyle\frac{1+\epsilon}{1-\epsilon}}G_E(Q^2) {\cal A}(Q^2,\epsilon)\right ]\right \}, \label{eq:eq7b} \\
{\cal N}^{(\mp)}P_z^{(\mp)})&=&\lambda_e\tau\sqrt{1-\epsilon^2}
\left [ |G_M(Q^2)|^2\mp 2 G_M(Q^2) Re {\cal A}(Q^2, \epsilon)\right. \nonumber\\
&&\left. \mp\sqrt{\displaystyle\frac{1+\tau}{\tau(1-\epsilon)}}G_M(Q^2)Re {\cal A}(Q^2, \epsilon)\right ],\nonumber
\end{eqnarray}
where $G_{E,M}^{(\mp)}(Q^2, \epsilon)=\mp G_{E,M}(Q^2)+\Delta G_{E,M}(Q^2, \epsilon)$. In (\ref{eq:eq7a}) and (\ref{eq:eq7b}) we used the fact that the FFs $G_{E,M}(Q^2)$ are real functions of $Q^2$.
The determination of the FFs is given by the measurement of these observables in $e^-N$ and $e^+N$-scattering, in identical kinematical conditions:
\begin{eqnarray}
\displaystyle\frac{d\sigma^{(-)}}{d\Omega_e}+\displaystyle\frac{d\sigma^{(+)}}{d\Omega_e}&=&
2\sigma_0\left [\epsilon G_E^2(Q^2)+\tau G_M^2(Q^2)\right ], \nonumber\\
\displaystyle\frac{1}{2}{\cal N}(P_x^{(+)}+P_x^{(-)})&=&-\lambda_e\sqrt{2\epsilon(1-\epsilon)\tau}
G_E(Q^2)G_M(Q^2), \label{eq:eq7c}\\
\displaystyle\frac{1}{2}{\cal N}(P_z^{(+)}+P_z^{(-)})&=& \lambda_e\tau\sqrt{(1-\epsilon^2)}
G_M^2(Q^2), \nonumber
\end{eqnarray}
Such experiment can not be presently realized, due to the absence of high quality polarized positron beams.

%%%%%%%%%%%%%%%%%%%%%%%%%%%%%%%%%%%%%%%%
\section{T-odd polarization observables}
%%%%%%%%%%%%%%%%%%%%%%%%%%%%%%%%%%%%%%%%%%%

We discuss here another possible strategy for the determination of the nucleon FFs, in absence of positron beam, i.e. using the measurement of different polarization observables in electron scattering, only. For this aim, the following set of T-odd observables must be measured:
\begin{itemize}
\item $P_y$ or $A_y$: the transversal component of the final nucleon polarization, in the collision of unpolarized leptons with unpolarized nucleon target, $e^-+N\to e^-+\vec N$ or the analyzing power, induced by the collision of unpolarized leptons (electrons or positrons) with a transversally polarized nucleon target $e^-+\vec N\to e^-+ N$;
\item the following components of the depolarization tensor, $D_{ab}$, $a,b=x,y,z$, which characterize the  dependence of the $b$-component of the final nucleon polarization on the $a$-
component of the  nucleon target, (in the scattering of longitudinally polarized leptons $\vec e^-+\vec N\to e^-+ \vec N$):
$$D_{xy}(\lambda_e),~D_{yx}(\lambda_e),~D_{yz}(\lambda_e),~
\mbox{and~}D_{zy}(\lambda_e),$$
\end{itemize}
The expressions for these polarization observables in terms of the nucleon electromagnetic FFs $G_{E,M}(Q^2)$ and for the $2\gamma$ amplitudes, ${\cal A}(Q^2,\epsilon)$ and $\Delta G_{E,M}(Q^2,\epsilon)$ can be written in the following form:
\begin{eqnarray}
{\cal N}P_y={\cal N}A_y&=&\sqrt{2\epsilon\tau(1+\epsilon)} {\cal I}_3(Q^2,\epsilon)+
\sqrt{{2\epsilon}(1-\epsilon)(1+\tau)} {\cal I}_1(Q^2,\epsilon),
 \nonumber\\
{\cal N}D_{xy}(\lambda_e)={\cal N}D_{yx}(\lambda_e)&=&2\lambda_e\epsilon
\sqrt{\tau(1+\tau)}{\cal I}_2(Q^2,\epsilon),\label{eq:eq9}\\
{\cal N}D_{yz}(\lambda_e)=-{\cal N}D_{zy}(\lambda_e)
&=&\lambda_e\sqrt{2\epsilon(1+\tau) (1+\epsilon)}{\cal I}_1(Q^2,\epsilon)
\nonumber
\end{eqnarray}
with the following notations:
$$
{\cal I}_1(Q^2,\epsilon)= G_E(Q^2) Im {\cal A}(Q^2,\epsilon),~
{\cal I}_2(Q^2,\epsilon)= G_M(Q^2) Im {\cal A}(Q^2,\epsilon),~
$$
\begin{equation}
{\cal I}_3(Q^2,\epsilon)=Im G_E(Q^2,\epsilon)G_M^*(Q^2,\epsilon)
\label{eq:eq16n}
\end{equation}
 
These quadratic combinations are uniquely determined solving Eqs. (\ref{eq:eq9}) as a set of three linear equations with respect to the quantities ${\cal I}_i(Q^2,\epsilon)$, i=1-3, at any values of both variables, $Q^2$ and $\epsilon$. The ratio 
$${\cal R}_{E,M}(Q^2)=
\displaystyle\frac{{\cal I}_1(Q^2,\epsilon)}{{\cal I}_2(Q^2,\epsilon)}
=\displaystyle\frac{G_E(Q^2)}{G_M(Q^2)}$$
has the most interesting  physical meaning, as it is related to the true FFs, real functions of $Q^2$, not to the 'generalized' FFs, function of two variables.

The three T-odd polarization observables enter in the determination of ${\cal R}_{E,M}(Q^2)$:
\begin{equation}
{\cal R}_{E,M}(Q^2)=- \displaystyle\frac{\lambda_e}{D_{xy}(\lambda_e)}\left [P_y+ \sqrt{\displaystyle\frac{1+\epsilon}{1-\epsilon}}
\displaystyle\frac{D_{zy}(\lambda_e)}{\lambda_e}\right ]
\sqrt{\displaystyle\frac{1-\epsilon}{2\epsilon}\tau}
\label{eq:eqx}
\end{equation}

This formula does not depend on the choice of the spin structure for the $2\gamma$ amplitude, it is correct for the tensor structure discussed in \cite{Re03b}, and for the axial structure as well.

Therefore, this relation indicates a possible way to extract the nucleon FFs with electron beams only, in presence of a $2\gamma$-contribution, without arbitrary assumptions.

We stress once more, that the formulas for all these observables are different for tensor and axial parametrizations, but Eq. (\ref{eq:eqx}), connecting directly the ratio $G_E(Q^2)/G_M(Q^2)$ to the observables has to be the same.

This method, without positron beam, looks more difficult, as all the T-odd polarization observables are small in absolute value, being of the order of $\alpha$. There are no experimental data, only upper limits for $P_y$ \cite{Ki70} and $A_y$ \cite{Po70}. No experimental attempt has been done to measure the components of the depolarization tensor. If the $2\gamma$ contribution is present and its contribution is sizeable, in absence of positron beams, polarization experiments will have to reach another level of precision. 

The number of independent T-odd observables is the same as the number of quadratic combinations $I_i(Q^2,\epsilon)$, $i=1-3$, Eq. (\ref{eq:eq16n}). 

Such polarization observable as ${\cal A}^{(e)}$, induced by the scattering of a transversally polarized electron beam ( by unpolarized nucleon target), being T-odd also, is not interesting at large momentum transfer, due to its proportionality to the small ratio $m_e/E$, where $E$ is the initial electron energy. 

Note that the observables $P_y$ and $D_{zy}$ allow us to determine ${\cal I}_3(Q^2,\epsilon)$ through the following relation:
\begin{equation}
{\cal I}_3(Q^2,\epsilon)=\displaystyle\frac{\cal N}{2\epsilon\sqrt{\epsilon\tau}}\left [P_y
\sqrt{1+\epsilon}+\sqrt{1-\epsilon}\displaystyle\frac{D_{zy}(\lambda_e)}{\lambda_e}\right ]
\label{eq:eq12}
\end{equation}
but there is no direct relation between this quantity and physical observables.

%%%%%%%%%%%%%%%%%%%%%%%%%%%%%%%%%%%%%%%%%%%%%%%%%%%%%%%%%%%
\section{T-even polarization observables}
%%%%%%%%%%%%%%%%%%%%%%%%%%%%%%%%%%%%%%%%%%%
We discuss in this section,  other polarization observables, $T-even$, which can also lead to the determination of the (real) nucleon FFs.

Writing the components of the depolarization tensor,  $D_{ab}$, $a,b=x,y,z$, for the scattering of unpolarized leptons, as:
$$ D_{ab}=D_{ab}^{(0)}+D_{ab}^{(1)},$$
(where $D_{ab}^{(0)}$ describes the main contributions to $D_{ab}$, which are independent from the axial amplitude ${\cal A}(Q^2,\epsilon)$, $D_{ab}^{(1)}$ describes the corresponding corrections, which are linear with respect to the amplitude ${\cal A}(Q^2,\epsilon)$) one can find for $D_{ab}^{(1)}$ the following expressions:
\begin{eqnarray}
{\cal N}D_{xx}^{(1)}&=&{\cal N}D_{yy}^{(1)}=0,\nonumber \\
{\cal N}D_{zz}^{(1)}&=&\sqrt{\tau(1+\tau)(1-\epsilon^2)}{\cal R}_M(Q^2,\epsilon),
\label{eq:eq13}\\
{\cal N}D_{xz}^{(1)}&=&-{\cal N}D_{zx}^{(1)}=-\sqrt{2\epsilon(1-\epsilon)(1+\tau)}
{\cal R}_E(Q^2,\epsilon),
\end{eqnarray}
where 
\begin{equation}
{\cal R}_{E,M}(Q^2,\epsilon)= G_{E,M}(Q^2)Re {\cal A}(Q^2,\epsilon)
\label{eq:eq14}
\end{equation}
We stress here once more that the formulas for $D_{ab}^{(1)}$ (in terms of the corresponding amplitudes) are different for axial and tensor parametrizations, but again, the physical answer for the ratio of electromagnetic FFs:
\begin{equation}
\displaystyle\frac{G_E(Q^2)}{G_M(Q^2)}= \displaystyle\frac
{{\cal R}_E(Q^2,\epsilon)}{{\cal R}_M(Q^2,\epsilon)}
\label{eq:eq15}
\end{equation}
in terms of T-even polarization observables has to be independent on the choice of the additional non-vector amplitude, either it is axial or tensor.

In the case of T-even observables, there is more freedom for the determination of five combinations of amplitudes, having seven possible observables:
\begin{equation}
\displaystyle\frac{d\sigma}{d\Omega_e},~P_x(\lambda_e),~P_z(\lambda_e),
D_{xx},~D_{yy},~D_{zz},~\mbox{and~}D_{xz},
\label{eq:eq16}
\end{equation}
The seven observables (\ref{eq:eq16}) are not independent, and are related by two model independent relations:
$$
D_{xx}+D_{yy}-D_{zz}=1,
$$
\begin{equation}
(1+\epsilon)^2 D_{xx}-(1-\epsilon)^2D_{yy}-4\epsilon D_{zz}
=2\epsilon\sqrt{1-\epsilon^2}\displaystyle\frac{P_z}{\lambda_e}
\label{eq:eq16a}
\end{equation}
at the level of the ($1\gamma+2\gamma$)--approximation.

The diagonal elements of the depolarization tensor $D_{ab}$ allow us to find the following combinations of amplitudes for $eN$ scattering:
\begin{eqnarray}
2\epsilon|{G}_E(Q^2,\epsilon)|^2&= &[D_{xx}+D_{yy}]{\cal N},\nonumber\\
2\epsilon\tau|{G}_M(Q^2,\epsilon)|^2&=& [-D_{xx}+D_{yy}]{\cal N},\label{eq:eq17}\\
\sqrt{\tau(1-\epsilon^2)(1+\tau)}G_M(Q^2)Re {\cal A}(Q^2,\epsilon)&=&\
1+\displaystyle\frac{1-\epsilon}{2\epsilon}
D_{xx}-\displaystyle\frac{1-\epsilon}{2\epsilon}
D_{yy}. 
\nonumber
\end{eqnarray}
Other two combinations, $Re G_E(Q^2,\epsilon)G_M^*(Q^2,\epsilon)$ and $G_E(Q^2) Re {\cal A}(Q^2,\epsilon)$ can be found through the observables $P_x(\lambda_e)$ and 
$D_{xz}$:
\begin{eqnarray}
{G}_E(Q^2) Re {\cal A}(Q^2,\epsilon)&=&- \displaystyle\frac{\cal N}
{2\epsilon\sqrt{2\epsilon(1+\tau)}}\left[
\displaystyle\frac{ P_x(\lambda_e)}{\lambda_e}
\sqrt{1+\epsilon}- 
D_{xz}\sqrt{1-\epsilon}\right ],\nonumber\\
Re G_E(Q^2)G_M^*(Q^2,\epsilon)&=&
\displaystyle\frac{\cal N}{2\epsilon\sqrt{2\epsilon\tau}}
\left [ \sqrt{1-\epsilon}\displaystyle\frac{P_x(\lambda_e)}{\lambda_e}-
D_{xz}\sqrt{1+\epsilon}\right ]
\label{eq:eq18}
\end{eqnarray}
Comparing Eqs. (\ref{eq:eq17}) and (\ref{eq:eq18}), one can find:
\begin{equation}
\displaystyle\frac{{G}_E(Q^2)}{{G}_M(Q^2)}=-\displaystyle\frac{1}{2\epsilon}
\sqrt{\tau\displaystyle\frac{1-\epsilon^2}{2\epsilon}}
\displaystyle\frac{
\left [ \sqrt{1+\epsilon}
\displaystyle\frac{P_x(\lambda_e)}{\lambda_e}-D_{xz}\sqrt{1-\epsilon}\right ]
}{
1+\displaystyle\frac{1-\epsilon}{2\epsilon}D_{xx}-
\displaystyle\frac{1+\epsilon}{2\epsilon}D_{yy}},
\label{eq:eq19}
\end{equation}
which involves five independent T-even polarization observables, in the same form as found for the tensor parametrization.

The key formulas, Eqs. (\ref{eq:eqx}) and (\ref{eq:eq19}) are the same for axial and tensor parametrizations, because both formulas describe the physical ratio ${G}_E^(Q^2)/{G}_M(Q^2)$. It is not the case for the generalized FFs. For example, in the tensor parametrization one finds:
\begin{eqnarray}
|{G}_E^{(t)}(Q^2,\epsilon)|^2&=&\displaystyle\frac{\cal N}{2\epsilon}
\left \{ 1+D_{zz}-\sqrt{ \displaystyle\frac{2\epsilon(1+\epsilon)}{\tau(1-\epsilon)}}
\left [ \sqrt{1+\epsilon}\displaystyle\frac{P_x(\lambda_e)}{\lambda_e}+ \sqrt{1-\epsilon}
D_{zx}\right ]\right \},\nonumber\\
Re {G}_E^{(t)}(Q^2,\epsilon) {G}_M^{(t)*}(Q^2,\epsilon)&=&
-\displaystyle\frac{\cal N}{\sqrt{2\epsilon(1-\epsilon)\tau}}
\left \{\displaystyle\frac{P_x(\lambda_e)}{\lambda_e}+(1-D_{yy})
\sqrt{2\tau(1-\epsilon)}\right \},
\label{eq:eq19a}
\end{eqnarray}
where the upper index $(t)$ indicates that these formulas are correct for the tensor parametrization.

These expressions differ essentially from the corresponding ones for axial parametrization, Eq.  (\ref{eq:eq17}). Of course, these formulas are equivalent at the level of the one-photon mechanism, as they depend on the choice of the parametrization of the $2\gamma$-contribution.

%%%%%%%%%%%%%%%%%%%%%%%%%%%%%%%%%%%%%%%
\section{Conclusions}
%%%%%%%%%%%%%%%%%%%%%%%%%%%%%%%%%%%%%%%
We analyzed the polarization effects in elastic lepton-nucleon scattering, using the axial parametrization of the two-photon contribution. The corresponding formulas are in general different from the case of tensor parametrization, which has been analyzed earlier \cite{Re03b}. However, the combinations of T-odd and T-even observables, which allow to determine the ratio $G_E(Q^2)/G_M(Q^2)$, without any assumption concerning the $2\gamma$ amplitudes, are the same for the two parametrizations. These formulas indicate a model independent way to measure the $G_E(Q^2)/G_M(Q^2)$ ratio, using an electron beam, only. It is, however, necessary to measure very small (of the order of $\alpha$) T-odd polarization observables, or to measure five T-even observables, with large accuracy.

The experimental evidence of the presence of $2\gamma$ contribution is mandatory, with a quantitative estimation of this effect as a function of $Q^2$, before involving such a complicated procedure for the measurement of nucleon electromagnetic FFs. If this effect appears in elastic $ep$ scattering already in the range of momentum transfer investigated at JLab, all findings based on the one-photon assumption will have to be reanalyzed at the light of new and complicated formalism. In this case, most of the advantages related to the electromagnetic probe would be lost, as it was indicated already long ago \cite{Gu73}.

{}
\end{document}